\documentclass[12pt,a4paper]{article}
\pdfoutput=1
\usepackage{jcappub}
\usepackage{bm}
\usepackage{bbm}
\usepackage{etoolbox}
\usepackage{mhchem}
\usepackage[english]{babel}
\usepackage{multirow}
\usepackage{url}
\usepackage{caption}
\usepackage{subcaption}
\usepackage[utf8]{inputenc}
\usepackage[dvipsnames]{xcolor}

\usepackage{graphicx}
\usepackage{xcolor}
\usepackage{amsmath,amssymb}
\usepackage{bm}
\usepackage{hyperref}
\usepackage{url}
\usepackage{color}
\usepackage{fleqn}
\usepackage{dcolumn}
\usepackage[normalem]{ulem}

\def\bea{\begin{eqnarray}}
\def\eea{\end{eqnarray}}
\def\be{\begin{equation}}
\def\ee{\end{equation}}
\newcommand{\gag}{g_{a\gamma}}

\newcommand{\Sec}[1]{Sec.~\ref{#1}}

\makeatletter
\gdef\@fpheader{}
\makeatother

\title{Bounds on Axions-Like Particles Shining in the Ultra-Violet}

\author[a,b, c]{Elisa Todarello} 
\emailAdd{elisa.todarello@nottingham.ac.uk}
\author[b,c]{and Marco Regis} 
\emailAdd{marco.regis@unito.it}

\affiliation[a]{School of Physics and Astronomy, University of Nottingham,
University Park, NG7 2RD, Nottingham, United Kingdom}
\affiliation[b]{Dipartimento di Fisica, Universit\`{a} di Torino, via P. Giuria 1, I--10125 Torino, Italy}
\affiliation[c]{Istituto Nazionale di Fisica Nucleare, Sezione di Torino, via P. Giuria 1, I--10125 Torino, Italy}

\abstract{
Axion-like particles (ALPs) can decay into two photons with a rest-frame frequency given by half of the ALP mass. This implies that ultra-violet searches can be used to investigate ALPs in the multi-eV mass range.
We use archival data from the Hubble Space Telescope between 110 and 170 nm to constrain ALPs with mass between 14.4-22.2 eV. We consider observations of a set of dwarf spheroidal galaxies and galaxy clusters and assume the ALP density in these objects to follow their dark matter density. The derived limit on the ALP-photon coupling $g_{a\gamma}$ excludes values above $10^{-12}~{\rm GeV}^{-1}$ over the whole mass range and surpasses previous limits by over one order of magnitude.
}

\begin{document}
\maketitle

\section{\label{sec:intro} Introduction}
Axion-like particles (ALPs) are compelling Dark Matter (DM) candidates, arising in various beyond-the-standard-model scenarios~\cite{Arias:2012az}.
As their cousin, the QCD axion~\cite{Peccei:1977hh,Peccei:1977ur, Weinberg:1977ma, Wilczek:1977pj}, they couple to photons through the characteristic vertex $\mathcal{L}=-\frac{1}{4}\gag\,a\,F_{\mu\nu}\tilde{F}_{\mu\nu},$ where $a$ is the ALP field, $F_{\mu\nu}$ is the electromagnetic field strength, $\tilde{F}_{\mu\nu}$ its dual, and $\gag$ the coupling constant.
Thanks to this interaction, an ALP can decay into two photons, each carrying an energy equal to half the ALP mass $m_a$ in the ALP rest frame. In astrophysical environments, dark matter ALPs have low momentum dispersion compared to their mass, implying that the signature of their decay into photons is a narrow spectral line.

Several searches have been conducted for such a spectral line in the optical~\cite{Grin:2006aw, Regis:2021, Todarello:2023hdk, Wang:2023imi}, and near-infra-red~\cite{Janish:2023kvi, Yin:2024lla} bands, yielding exclusion limits on $\gag$ surpassing those from stellar evolution in globular clusters~\cite{Ayala:2014pea, Dolan:2022kul}. Until now, no such search had been performed in the Far Ultra-Violet (FUV).
A different strategy consists of directly or indirectly looking at the continuum cosmological ALP signal given by the cumulative UV emission from the collection of lines arising in different DM halos~\cite{Bernal:2022xyi,Bernal:2022wsu,Nakayama:2022jza,Carenza:2023qxh,Porras-Bedmar:2024uql,Libanore:2024hmq}.

In this work, we use archival spectroscopic data from the Hubble Space Telescope (HST) to put the most stringent limits to date on DM ALPs in the mass range 14.4-22.2~eV, by searching for excess radiation due to ALP decay into photons in individual halos. Ideal targets for such a search are dark-matter-rich environments such as dwarf spheroidal galaxies (dSphs) and galaxy clusters. We consider HST observations of four promising targets: two dwarf spheroidals, Ursa Minor (UMi) and Draco, and two galaxy clusters, Virgo, and Fornax. Although these observations were not thought for a DM search, and bright objects are present in the field of view, we are able to constrain the axion to photon coupling to values $\lesssim  10^{-12}$~GeV over the whole mass range, reaching $\gag < 2\times 10^{-13}$~GeV for some masses. Previous limits in this mass range \cite{Carenza:2023qxh, Porras-Bedmar:2024uql} are thus improved by more than one order of magnitude on average.

The axion decay rate scales as $m_a^3$, and one could expect a stronger bound in the FUV compared to those in the optical band, for example from MUSE~\cite{Todarello:2023hdk}. However, in addition to the presence of bright objects, the constraining power of our current search is limited by three main factors, one inevitable, and two due to the observational setup. The inevitable factor is extinction due to dust particles. At the frequencies considered here, extinction due to galactic dust along the lines of sight to our targets causes an average flux reduction ranging from 20\% to 55\%. In addition, extinction internal to the target has to be taken into account. The second limiting factor is the relatively poor spectral resolution of the observations, especially when targeting extended sources like diffuse emission from a DM halo. Third, the field of view of the observations is very narrow $\sim 10-20$ arcsec$^2$. 
This highlights the importance, in the context of DM ALP searches, of wide-field integral field units (IFU) observations for future UV satellites, since they can enable simultaneous spatial and spectral mapping of diffuse emissions.

The structure of this paper is as follows. The data from HST observations are presented in~\Sec{sec:data}. The calculation of the ALP decay signal is described in~\Sec{sec:axion}. The statistical analysis and results are discussed in~\Sec{sec:res}. We conclude in~\Sec{sec:conc}.

\section{\label{sec:data} Data}
\begin{table}
\scriptsize
    \centering
    \begin{tabular}{|c|c|c|c|c|c|c|c|c|c|}
        \hline 
       Target & Obs. ID & $\ell$  & $b$  & $D$  & Aperture & Ang. sep. &   Min. dist. & $r_s$ \\ 
        &  & [deg] &  [deg] &  [$10^{21}$~eV~cm$^{-2}$]  & [arcsec$^2$] &  [arcmin] & [kpc] & [kpc]\\\hline \hline 
        UMi & O56N01010  & 105.08 & 44.78 & $4.2\pm 0.5$ &  52x0.5 & 5.180 & 0.115 & 0.61\\ \hline 
        Draco & O5GS24010 &  86.37 & 34.72 & $5.4\pm 0.2$ &  52x0.5 & 5.632 & 0.124 & 0.35\\ \hline 
        Virgo & O56004020 & 283.78 & 74.49 & $93\pm7$ ($119\pm 10$) &   52x0.5 & 0.522 & 2.460 & 380 \\ \hline 
        Fornax & O4AK04010-4040 & 236.72 & -53.64 & $300\pm13$ ($295\pm22$)  &  52x2 & 0.126 & 0.725 &131\\ \hline 
    \end{tabular}
    \caption{Observations used in this work. The columns are: target name, HST observation ID, galactic longitude of the observation, galactic latitude of observation, D-factor over the solid angle corresponding to about half the nominal aperture (as appropriate for MAMA observations), observation aperture, angular separation from the center of the object in arcminutes, minimum distance of the line of sight from the center of the object, and reference scale radius of the DM halo assuming an NFW profile. The center of the object is taken from the Simbad database~\cite{Wenger:2000sw}. All observations were obtained with the G140L grating.}
    \label{tab:data}
\end{table}

\begin{figure}[t!]
\centering
   \includegraphics[width=0.7\textwidth]{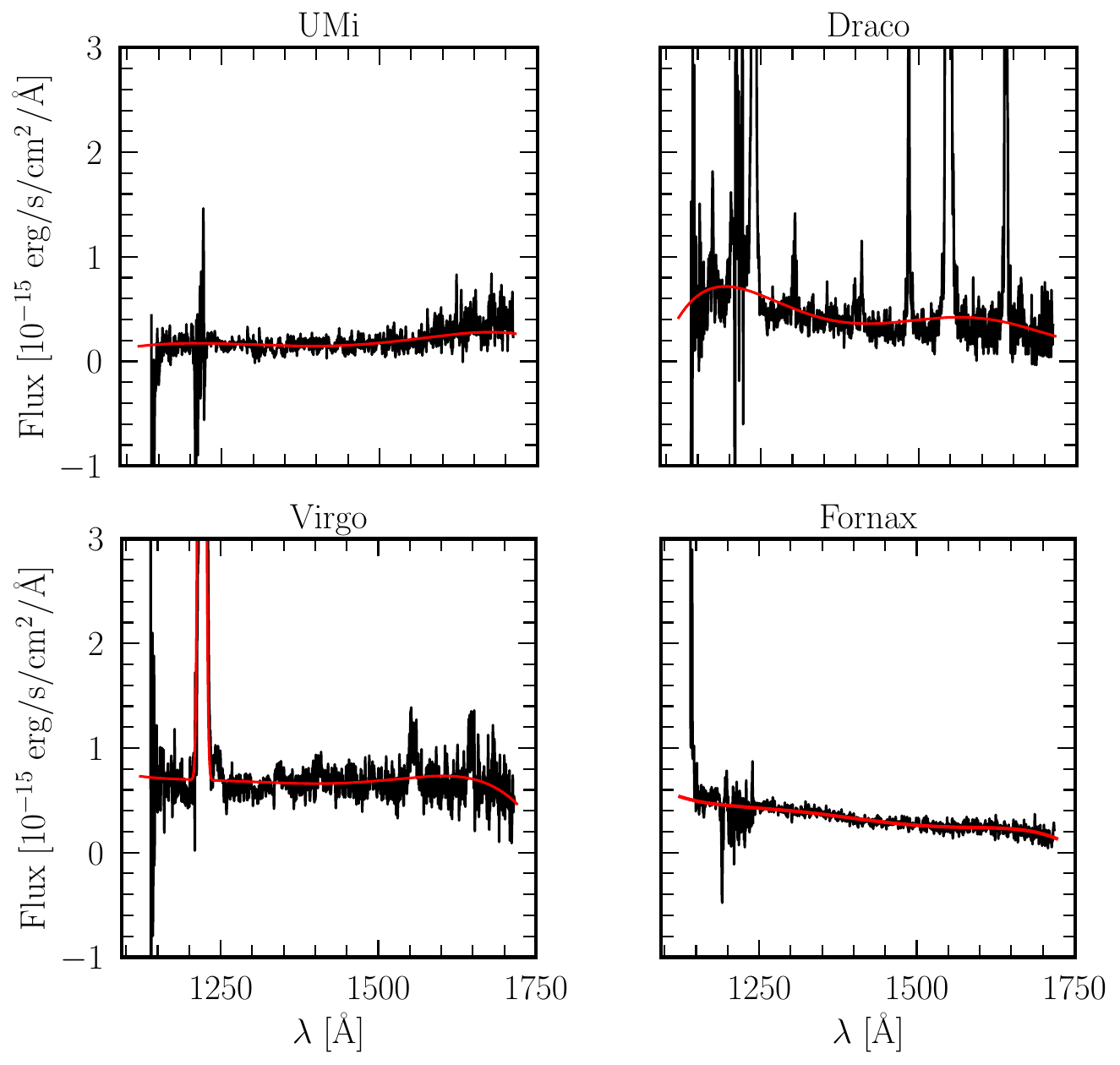}
    \caption{Spectra used in our analysis (black lines). For Fornax, we show the spectrum averages over all corresponding datasets. The red lines represent the 5th-order polynomial we use to fit the continuum emission and subtract from the spectra, plus a Gaussian fit to the Ly$\alpha$ line for Virgo. See the main text for more details.
}
\label{fig:data}
 \end{figure}

We use archival HST data collected with the Space Telescope Imaging Spectrograph Instrument (STIS) FUV Multi-Anode Microchannel Array (MAMA) detector~\cite{stis}. All data are long slit spectra obtained through the first-order G140L grating, covering the spectral range 1150–1730~\AA\ and processed through the STIS calibration pipeline\footnote{\href{https://hst-docs.stsci.edu/stisdhb/chapter-3-stis-calibration}{https://hst-docs.stsci.edu/stisdhb/chapter-3-stis-calibration}}. The datasets are listed in Table~\ref{tab:data}. We analyzed one observation of each of the targets UMi, Draco, and Virgo, and four observations of Fornax.
We also found observations of the A2975 galaxy cluster, which are in principle promising, but given the lower quality of the data, compared to the other targets, were excluded from our analysis. The slits used have a length of $52''$ and a width of $0.5''$ (UMi, Draco, Virgo) or $2''$ (Fornax). The length of the slits, as projected onto the MAMA detectors, is $25''$.
Thus we will integrate the ALP signal over a spatial region of $25\times 0.5''$ (UMi, Draco, Virgo) and  $25\times 2''$ (Fornax).
 
Different slits correspond to different spectral resolutions. For extended sources that fill up the slit, like the diffuse emission from the DM halo we're searching for here, the FWHM spectral resolution for MAMA first-order modes with a 52X0.5 slit is about 20 bins across the whole spectral range. Combined with the average dispersion of the G140L grating of 0.6~\AA/pixel, it yields a FWHM resolution in the range $0.7\textrm{--}1\times 10^{-2}$.
The 52x2 slit used for the Fornax observation has instead a FWHM spectral resolution of about 80 bins.

The data are shown in Figure~\ref{fig:data}. For Fornax, we show the spectrum averaged over the four datasets. 
In each panel, we can notice a continuum emission. In particular, the observations are pointed at the following objects: the QSO J1508+6717 quasar (UMi), the Draco-C1 symbiotic binary star system (Draco), the M87 Active Galaxy Nucleus (Virgo), and the Brightest Galaxy in a Cluster (BCG) NGC 1399 (Fornax). A smooth background subtraction is already performed at the calibration level but with continuum background estimated from a faint region away from the source. We thus subtract the continuum emission from the source region. To derive the continuum, we first remove fluxes larger than $N_\sigma$ standard deviation above the mean flux. We choose $N_\sigma$ in such a way as to maximize the goodness of the polynomial fit to be performed. After removing these large values, we calculate again the average flux over all bins. This new average is used to replace the large values that were previously identified, as well as any negative flux value. We then smooth the spectrum by convolving it with a Gaussian window function with a standard deviation equal to five bin widths. Finally, we fit a fifth order polynomial to the smoothed spectrum. The result of this procedure is shown by the red lines in Fig.~\ref{fig:data}.  In the case of multiple observations of Fornax, we determine the continuum spectrum on each dataset separately.

In addition to the continuum, several emission spectral lines are visible. From the STIS calibration pipeline, the only two lines that are subtracted are the Ly$\alpha$ and O I (1300~\AA) lines, again estimated from a fainter region. In particular, in our source region, additional contamination from the Ly$\alpha$ line at 1215.7~\AA~is visible in all data sets and will negatively affect our bounds. In the case of Virgo, the line residual is very bright and we are able to model it as a Gaussian centered at a wavelength of 1.221~\AA\ with a standard deviation of 4~\AA, of order the spectral resolution for extended sources. For the other three targets, the residuals from the subtraction of the Ly$\alpha$ line are rather irregular and a simple Gaussian fit do not provide a good description of data.
Conservatively, we do not apply further subtractions related to the Ly$\alpha$ line in targets other than Virgo, nor to other lines that are present in the spectrum (e.g., very visible in the Draco case).
The red lines in Fig.~\ref{fig:data} mark our best-fit continuum model (plus Ly$\alpha$ model in Virgo).


\section{\label{sec:axion} ALP signal}

The flux density at wavelength $\lambda$ produced by decays of ALPs from a given direction in the sky, identified by $\theta$, can be computed as:
\be
S_\lambda (\theta)=\frac{\Gamma_a}{4\pi}\,\frac{1}{\sqrt{2\pi}\sigma_\lambda} \exp{\left[-\frac{(\lambda-\lambda_{obs})^2}{2\sigma_\lambda^2}\right]}\, \frac{D}{m_a}\,2\,E_{obs}\ e^{-\tau_\lambda}\;,
\label{eq:flux}
\ee
where $\Gamma_a$ is the ALP decay rate, $\lambda_{obs}=4\pi\,c\,(1+z)/m_a$ is the wavelength of observation, $\sigma_\lambda$ is the HST spectral resolution, $E_{obs}=m_a/2/(1+z)$ is the energy of the observed photons, the factor of $2$ accounts for the two photons produced in each decay, $m_a$ is the ALP mass, and $\tau_\lambda$ is the optical depth. The D-factor is given by
\be
D = \int_{\Delta\Omega}  d\Omega\, \int d\ell\, \rho_a[r(\theta,\Omega,\ell)]\;,
\ee
with $\Omega$ being the solid angle and $\Delta \Omega$ referring to the observed one. The DM spatial profile $\rho_a$ is a function of the radial distance from the center of the DM halo $r$, which can be expressed in terms of the luminosity distance along the line of sight $\ell$ and the direction of observation $\theta$.
The decay rate $\Gamma_a$ depends on the ALP mass $m_a$ and the effective ALP-two-photon coupling $\gag$ through
$\Gamma_a=\gag^2\,m_a^3/(64\pi)$.
We assume a Gaussian behavior for the energy response of the detector, with FWHM described in Section~\ref{sec:data}, corresponding to $\sigma_\lambda \sim 5.1$~\AA~for all targets except Fornax, for which $\sigma_\lambda \sim 20.4$~\AA.

In Eq.~\ref{eq:flux}, we neglect the velocity dispersion of ALPs in the DM halo since it is smaller than the spectral resolution. 
This is certainly the case in dwarf galaxies where it is typically $\sim 10^{-5}$, and also true for clusters since their velocity dispersion is $\lesssim 10^{-2}$, i.e.,  comparable but smaller than the HST spectral resolution considered here.
We include the wavelength shift arising from the dwarfs' heliocentric radial velocity, which is however negligible, as well as from the redshift in the Virgo ($z=0.0038$), and Fornax ($z=0.0046$), 
where it leads to a minor but not irrelevant correction.

Inserting units to estimate Eq.~\ref{eq:flux}, one obtains:
\be
S_\lambda (\theta)\simeq 32\,\mu{\rm Jy}\,\left(\frac{\gag}{10^{-11}\,{\rm GeV}^{-1}} \right)^2 \left(\frac{m_a}{20\,{\rm eV}} \right)^2\, \left(\frac{D}{10^{22}\,{\rm eV/cm}^2} \right)\,\left(\frac{10^{-2}}{\sigma_\lambda/\lambda} \right)\,\frac{e^{-\tau_\lambda}}{1+z}\;.
\label{eq:fluxunits}
\ee

Under the assumption of spherical symmetry, the DM density $\rho_a(r)$ is a function only of the radial distance $r$.
Our reference scenario for the description of the DM density profile is given by the NFW functional form~\cite{Navarro:1995iw}:
\be
\rho_{\rm{NFW}}(r)=\frac{\rho_s}{\left(\frac{r}{r_s}\right)\left( 1 + \frac{r}{r_s} \right)^2}\;,
\label{eq:rho}
\ee
where $\rho_s$ and $r_s$ are, respectively, the scale density and radius.

In the case of clusters, these two parameters are derived from the mass and concentration of each target. More concretely, we take $M_{200}=4.2\pm0.5\,\times 10^{14}\,M_{\odot}$ for Virgo~\cite{McLaughlin:1998sb}, and $M_{200}=5.1^{+0.6}_{-0.5}\,\times 10^{13}\,M_{\odot}$ for Fornax~\cite{Schellenberger:2017wdw}.
The concentration parameter is taken from Ref.~\cite{Ishiyama:2020vao}. To determine the central value (uncertainty) in the D-factor we take the best-fit ($2\sigma$ limits) of the above mass estimates, derive the associated concentration~\cite{Ishiyama:2020vao}, and compute the D-factor. Values are reported in Table~\ref{tab:data}. The statistical uncertainty on the concentration of Ref.~\cite{Ishiyama:2020vao} is negligible, but to take into account a possible systematic uncertainty due to modeling, we consider a different mass-concentration relation, from Ref.~\cite{Sanchez-Conde:2013yxa} (associated D-factor values are within brackets in Table~\ref{tab:data}). We will show our bounds as bands, combining the two different mass-concentration relations.
For what concerns the dSph galaxies, we derive the D-factor and associated uncertainties using the NFW parameters obtained from Jeans analyses in Draco~\cite{Regis:2023rpm} and Ursa Minor~\cite{Ullio:2016kvy}. 

The shape of the DM profile at the center of dSph galaxies is not fully understood. In order the test the impact of a central core instead of the NFW cusp, we consider the Burkert profile derived in Ref.~\cite{Regis:2023rpm} (Draco) and  Ref.~\cite{Ullio:2016kvy} (UMi). We obtain $D=5.6\pm0.1\times 10^{21}$~eV~cm$^{-2}$ (Draco) and  $D=4.1\pm0.5\times 10^{21}$~eV~cm$^{-2}$ (UMi), i.e., negligible differences with respect to the NFW case.
This is understandable, since the observations considered in this work are quite away from the center, about 5 arcmin, and at a radius where the measured dispersion velocity of stars constrains well the mass inside that radius, which means that different DM profiles must lead to a similar DM density. 
In the case of clusters of galaxies, cored profiles are disfavored. Our assessment of the uncertainty on the DM density thus relies on estimating the uncertainties in the parameters of the NFW profile, as discussed above.

\begin{figure}[h]
    \centering
    \includegraphics[width=0.7\textwidth]{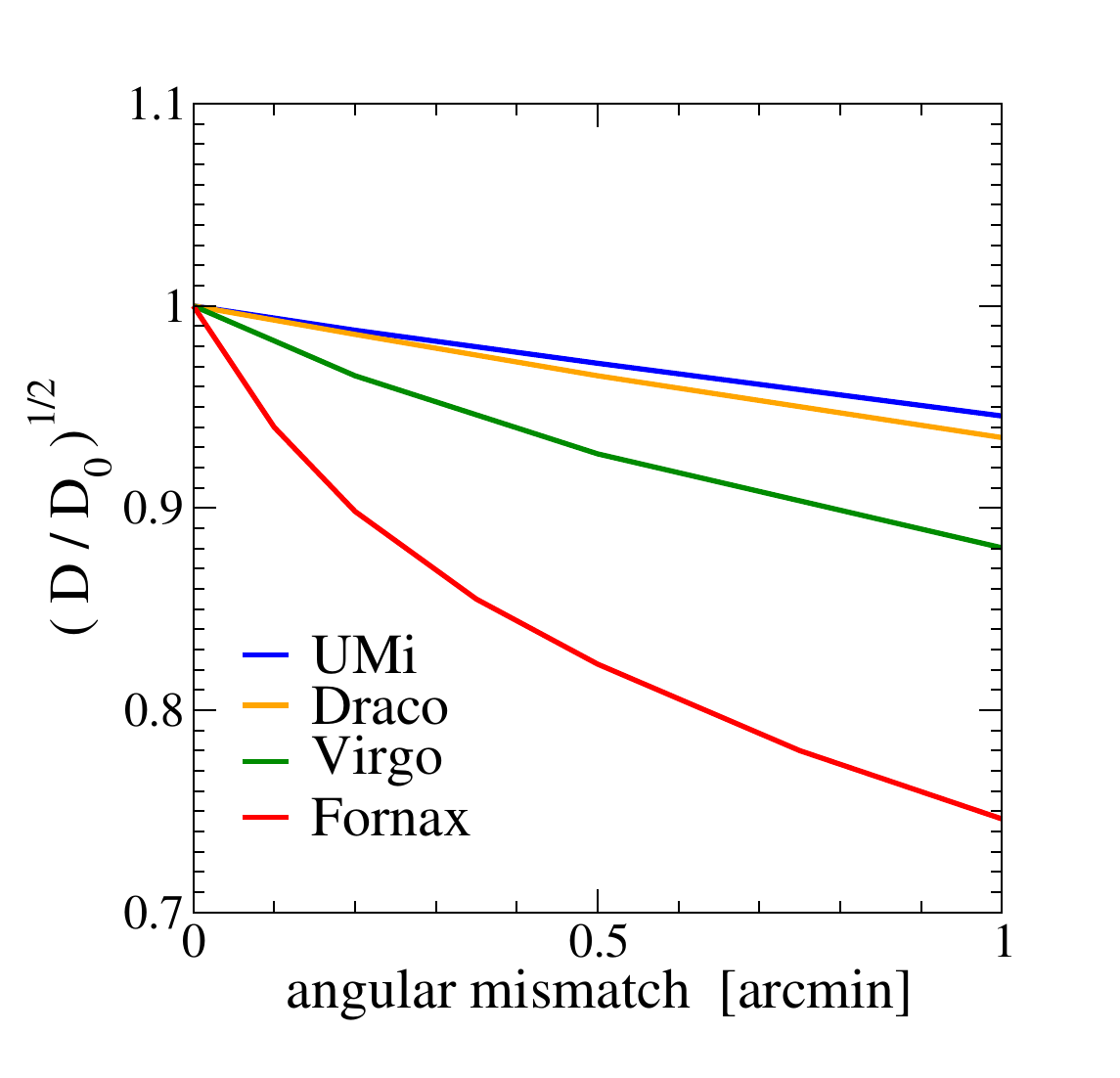}
    \centering
    \caption{Reduction of $\sqrt{D}$ with respect to our reference case of Table~\ref{tab:data} by considering the DM center offset from the target center and away from the direction of observation by an angular distance provided in the x-axis. The quantity $\sqrt{D/D_0}$ gives the impact on the $g_{a\gamma}$ bounds.
    }
    \label{fig:off}
\end{figure}

In our computation, we consider the slit size (about half of the nominal, see Section~\ref{sec:data}) and orientation of the observations, with a beam function equal to one inside the slit and zero outside, with sharp boundaries. We also assume that the center of the DM halo coincides with the optical (for dSphs) and X-ray (for clusters) center of the target.
To test the impact of this assumption, we show in Fig.~\ref{fig:off} how the bounds would weaken if the DM center is offset from the assumed center, and taking the opposite direction with respect to the observations (i.e., the most pessimistic case). We show the ratio of $\sqrt{D}$ since this is the factor entering in the determination of the $g_{a\gamma}$ bound.
In the case of dSph, the effect is marginal, essentially because data are not from the central region of the object.
In the case of clusters, and in particular Fornax, it could be more relevant, up to a few tens of percent, for an arcmin (a few kpc) offset. Clearly if the offset would be such to make the DM center closer to observations, bounds on $g_{a\gamma}$ would instead become stronger. It is therefore hard to properly quantify this uncertainty without a dedicated study assessing the likelihood and direction of an offset, something beyond the goal of the present work. Given the relatively small impact on the $g_{a\gamma}$ bounds, we refer the reader to Fig.~\ref{fig:off} for an estimate of the maximal change.

Dust extinction plays an important role at UV frequencies. The optical depth in Eq.~\ref{eq:flux} is related to the extinction in magnitudes as $\tau_\lambda = 0.92 A_\lambda$.
We consider three sources of extinction: the Milky Way, the intracluster medium along the line of sight (only for clusters),  and the galaxy targeted by the observations.

We extract $A_\lambda$ due to the extinction along the line of sight in the Milky Way from Ref.~\cite{extinct}.
We adopt the extinction law from Ref.~\cite{1989ApJ...345..245C}, checking that with an alternative model from Ref.~\cite{1994ApJ...422..158O} results are nearly identical. All targets are off the galactic plane, so the extinction is moderate and the model prediction is robust.

In the case of dSph galaxies, there are no other sources of extinction since their dust content is extremely low~\cite{2012AJ....144....4M}.
The picture for clusters can be different.
Concerning the intracluster medium, we follow Ref.~\cite{virgodust}, where a measurement of $A_V=0.14$ in Virgo was reported. In Fornax, there are no measurements, and we adopt the same value of Virgo, just rescaled under the assumption that the ratio of the gas mass between the two clusters follows the ratio of their total mass.
We note that lower values $A_V\ll 0.14$ are more typical in clusters~\cite{Gutierrez:2014wfa,Shchekinov:2022rpm}, so the latter choice should be conservative. 
The dust content of the BCG in Virgo (M87) and Fornax (NGC 1399) is low, thus their intrinsic extinction is negligible with respect to the other contributions, and we disregard the extinction from the BCGs in our computation.

With the description outlined above, there are two free parameters remaining from Eq.~\ref{eq:flux}, $\gag$ and $m_a$, that will be sampled in our scans.

\section{ Methods and Results}\label{sec:res}
Our statistical analysis is based on the comparison between the expected ALP signal and the observed spectra by means of a Gaussian likelihood
\be 
\mathcal{L} = e^{-\chi^2/2} \;\;\; 
{\rm with} \;\;\; 
\chi^2 = \sum_{i=1}^{N_{bin}}\left(\frac{S_{th}^i-S_{obs}^i}{\sigma_i}\right)^2\;,
\label{eq:like}
\ee
where $i$ runs over the spectral HST bins, $S_{th}^i$ is the theoretical estimate for the flux density, $S_{obs}^i$ is the observed flux density, and $\sigma_i$ is the statistical error. Before computing the $\chi^2$, we subtract the continuum, and Ly$\alpha$ in the case of Virgo, from the data. 
For HST MAMA observations, the raw data statistical error in each pixel is given by the square root of the observed number counts\footnote{\href{https://hst-docs.stsci.edu/stisdhb/chapter-2-stis-data-structure/2-5-error-and-data-quality-array} {https://hst-docs.stsci.edu/stisdhb/chapter-2-stis-data-structure/2-5-error-and-data-quality-array}}, and we use the flux errors provided with data, obtained by propagating the error on raw data through the calibration pipeline. No correlation among different energy bins is available, and we will assume it to be negligible in our analysis.
Bins flagged for bad data quality are excluded from the computation.

We assume that $\lambda_c(\gag)=-2\ln[\mathcal{L}(g_{a\gamma})/\mathcal{L}(\gag^{b.f.})]$ follows a $\chi^2$-distribution with one d.o.f. and one-sided probability given by $P=\int^{\infty}_{\sqrt{\lambda_c}}d\chi\,e^{-\chi^2/2}/\sqrt{2\,\pi}$, where $\gag^{b.f.}$ denotes the best-fit value for the coupling at a specific ALP mass. 
For a given ALP mass $m_a$, we thus derive the 95\% C.L. bound on $\gag$, by requiring $\chi^2 =\chi^2_{b.f.} +2.71$, where $\chi^2_{b.f.}$ is the $\chi^2$ of the best-fit model.

\begin{figure}[t!]
\centering
   \includegraphics[width=\textwidth]{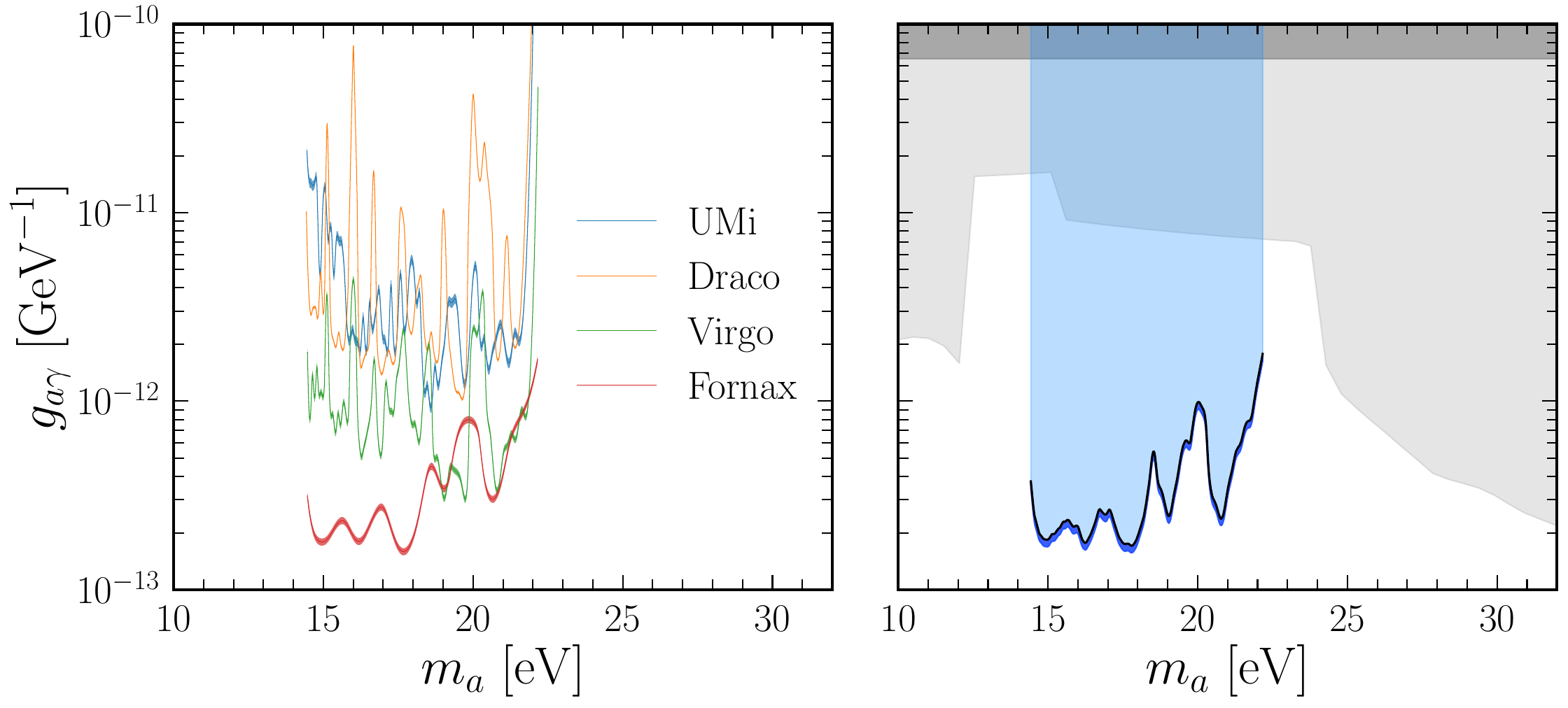}
    \caption{95\% C.L. upper limits on the effective ALP-two-photon coupling $\gag$ as a function of the ALP mass, for each target listed in Table~\ref{tab:data} individually (left) and combined (right). On both panels, the thickness of the lines represent the uncertainty associated with the D-factor (see Table~\ref{tab:data}). The solid black line marks our most conservative exclusion limit. On the right panel, the region shaded in dark gray is excluded by stellar evolution in globular clusters~\cite{Ayala:2014pea}, while the region in light gray is excluded by Refs.~\cite{Carenza:2023qxh,Porras-Bedmar:2024uql,Wang:2023imi, Wadekar:2021qae, Bolliet:2020ofj}, assuming ALPs are the dark matter. \
}
\label{fig:bounds}
 \end{figure}

The values of the parameters describing the polynomial fit are not altered by the addition of the ALP line signal. This is because the latter affects a small fraction of the spectral bins and thus contributes very mildly to the fit of the entire spectrum.
To compute the ALP theoretical signal from Eq.~\ref{eq:flux}, we need to model the D-factor. In order to bracket the associated systematic uncertainty, we defined two limiting scenarios, with high and low estimates of $D$, as discussed in the previous Section. The thickness of the bounds in Fig.~\ref{fig:bounds} encloses all the $\gag$ limits obtained assuming $D$ in between the highest upper limit and the lowest lower limit between the ones reported in Table~\ref{tab:data}.

The strength of the bounds depends on the size of the D-factor and on the quality of the data. In the case of dSphs galaxies, we found lower values of $D$ with respect to clusters, see Table~\ref{tab:data}, essentially because observations are more distant from the center of the object (i.e., of the DM halo). This leads to weaker bounds on ALPs in dSphs, see the left panel of Fig.~\ref{fig:bounds}.
Fornax is the most constraining target in nearly the whole mass range because of the better statistics (four observations available), lower extinction than Virgo and larger D-factor, due to a larger aperture and pointings closer to the center. On the other hand, it has a poorer spectral resolution than other targets, and this especially impacts bounds from the small wavelength range, where the spectrum is contaminated by other emissions.

Finally, we combine all 4 targets together by treating them as independent tests, i.e., writing the global likelihood as the product of the individual ones
\be
 \mathcal{L}^{all}(\gag)=\prod_{j=1}^4 \mathcal{L}^j(\gag)\quad{\rm i.e.,}\quad \chi^2_{all}(\gag)=\sum_{j=1}^4 \chi^2_j(\gag)\;.
\label{eq:likecomb}
\ee
The corresponding bound is shown in the right panel of Fig.~\ref{fig:bounds}.

As expected, the combined bound is dominated by Fornax.
We can notice that, for some masses, the combination leads to a deterioration, rather than an improvement, of the bound with respect to the best bound one could infer from the different curves in the left panel. This is due to the presence of unsubtracted astrophysical lines in some targets, and it is particularly pronounced at around 20.5~eV due to the contamination from the Ly$\alpha$ line, which has been only partially subtracted in our analysis.

\section{\label{sec:conc} Conclusions}

In this work we have constrained the photon-ALP coupling $\gag$ for ALP masses between 14.4 and 22.2, see Fig.~\ref{fig:bounds}. 

ALP models generically include an explicit or effective coupling between ALPs and electromagnetism. This opens the possibility for ALPs to decay into two photons. Assuming ALPs to constitute the cold DM component of the Universe, the decay signal is a nearly monochromatic line coming from DM halos and with a rest-frame frequency given by half of the ALP mass.

Dwarf spheroidal galaxies have very large mass-to-light ratios, and clusters are the most massive bound systems in the Universe. Therefore they are both prime targets in indirect DM searches.
We have analyzed a selection of dSphs (Draco and Ursa Minor) and clusters (Virgo and Fornax) searching for a signal from ALP decay.

To investigate masses between 14.5 and 21.9 eV, we used ultraviolet observations from the HST between 110 and 170 nm.  
Although HST is not optimized to conduct spectroscopic searches of spatially diffuse emissions, still its observations have the capability to set relevant constraints on ALPs. 

Our analysis constrains the photon-ALP coupling $\gag$ to a level below  $10^{-12}\,{\rm GeV}^{-1}$. 
The bounds are reported in Fig.~\ref{fig:bounds} and compared to existing ones from the literature. The limits from our work are significantly more constraining than current limits from other probes, in the same mass range. 
Our analysis includes an estimate of the uncertainty associated to the determination of the DM density in the observed targets.

Future telescopes and observations will be able to pursue the strategy presented here with higher sensitivity, to further test the ALP hypothesis.
The Xuntian telescope~\cite{CSST:2021} aims to dramatically improve the field of view with respect to HST, offering the possibility to significantly enhance the D-factor. The UVEX~\cite{Kulkarni:2021tit} telescope will perform an all-sky survey allowing us to combine many different targets.
On a shorter time scale, dedicated observations with the HST, such as towards the center of dSph galaxies with an exposure time significantly larger than  10~ks, can contribute to push downwards the $\gag$ limits.

\section*{Acknowledgements}
ET is grateful to Alfonso Aragon-Salamanca and Nina Hatch for useful discussions.
ET is supported by STFC Consolidated Grant [ST/T000732/1].
MR and ET acknowledge support from the project ``Theoretical Astroparticle Physics (TAsP)'' funded by the INFN. This article/publication is based upon work from COST Action COSMIC WISPers CA21106, supported by COST (European Cooperation in Science and Technology)”. 

\textbf{Data Availability Statement}: The data used in this work is publicly available from the Mikulski Archive for Space Telescopes (MAST)~\cite{mast}.

\bibliographystyle{JHEP_improved}
\bibliography{biblio}

\end{document}